\documentclass[a4paper,fleqn,usenatbib]{mnras}

\usepackage{amsopn}

\usepackage{color}
\usepackage{caption}
\usepackage{relsize}
\usepackage{graphicx}
\usepackage{amsmath}
\usepackage{amssymb}
\usepackage{pdflscape}
\usepackage{rotating}
\usepackage{longtable}
\usepackage{afterpage}

\title[Multiwavelength correlations and flares in blazars]{Multiwavelength Cross-Correlations and Flaring Activity in Bright Blazars}
\author[Liodakis et al.]
{I. Liodakis$^{1}$\thanks{ilioda@stanford.edu}, R. W. Romani$^{1}$, A. V. Filippenko$^{2,3}$, S. Kiehlmann$^{4}$, W. Max-Moerbeck$^{5}$,\newauthor A. C. S. Readhead$^{4}$, and W. Zheng$^{2}$ \\
$^{1}$KIPAC, Stanford University, 452 Lomita Mall, Stanford, CA 94305, USA\\
$^{2}$Department of Astronomy, University of California, Berkeley, CA 94720-3411, USA\\
$^{3}$Miller Senior Fellow, Miller Institute for Basic Research in Science, University of California, Berkeley, CA  94720, USA\\
 $^{4}$Owens Valley Radio Observatory, California Institute of Technology, Pasadena, CA 91125, USA\\
$^{5}$Universidad de Chile, Departamento de Astronom\' ia, Camino El Observatorio 1515, Las Condes, Santiago, Chile\\
}

\begin{document}

\maketitle
\label{firstpage}
\begin{abstract}
Blazars are known for their energetic multiwavelength flares from radio wavelengths to high-energy 
$\gamma$-rays. In this work, we study radio, optical, and $\gamma$-ray light curves
of 145 bright blazars spanning up to 8~yr, to probe the flaring activity and
interband correlations. Of these, 105 show $>1\sigma$ correlations between one or more wavebands, 26 of which have a $>3\sigma$ correlation in at least one wavelength pair, as measured by the discrete correlation function. The most common and strongest correlations are found between the optical and $\gamma$-ray bands, with fluctuations simultaneous within our $\sim 30$~d resolution. The radio response is usually substantially delayed with respect to the other wavelengths with median time lags of $\sim 100$--160~d. A systematic flare identification via Bayesian block
analysis provides us with a first uniform sample of flares in the three bands, allowing
us to characterise the relative rates of multiband and ``orphan'' flares. Multiband flares tend to have higher amplitudes than ``orphan'' flares.
\end{abstract}


\section{Introduction}\label{sec:intro}

Blazars are luminous and highly variable across the entire electromagnetic spectrum. Their emission, dominated 
by relativistically boosted jets, shows a spectral energy distribution (SED) with two distinct humps.
The low-energy peak (radio to ultraviolet, and in some cases X-rays) is thought to be produced by synchrotron radiation, and the 
high-energy hump (X-rays to high-energy $\gamma$-rays) is likely Compton emission. However, the source of the seed photons that produce the observed high-energy emission is not fully understood. Some models also attribute $\gamma$-ray flux to proton-mediated emission rather than inverse-Compton scattering. Traditionally, blazars are classified  by their optical spectral line properties as 
BL Lacertae objects (BL Lacs; broad lines typically lost against a bright continuum) and flat-spectrum radio quasars (FSRQs; strong, broad lines).

Blazars exhibit violent flaring, in which the Earth-directed flux can increase by orders of magnitude over short timescales. 
Flares appear in multiple wavebands either simultaneously or with a time delay ranging from days to months. 
These delays can provide clues to the emission process and the relative location of the emitting region at different 
wavelengths. Flaring events are often accompanied by other phenomena such as the ejection of jet components 
mapped at very long baseline interferometry (VLBI) scales (e.g., \citealp{Marscher2008}) and rotations of the optical polarization plane 
(e.g., \citealp{Blinov2017}). The origin of the $\gamma$-rays in particular is poorly understood.
The most likely mechanism appears to be inverse-Compton (IC) scattering of lower-energy photons by the 
relativistic jet electrons. However, the nature of the incident photon field and the location of the upscattering 
electrons are poorly constrained. If the incident photon field is external to the jet (accretion disk, 
broad-line region, etc.), the IC scattering is referred to as external Compton (EC, e.g., \citealp{Dermer1992}), while if the incident photons 
are from the jet itself it is called synchrotron self-Compton (SSC, e.g., \citealp{Abdo2010-II}). One can probe these uncertainties by studying the correlated variability between different frequency bands.

Several monitoring programs have followed $\gamma$-ray-loud blazars at various wavelengths since the launch of the {\it Fermi} 
gamma-ray space telescope, with the primary goal of constraining the mechanism and location of the $\gamma$-ray emission through time-series analysis. This is usually accomplished through the cross-correlation of radio and $\gamma$-ray light curves (e.g., \citealp{Fuhrmann2014,Max-Moerbeck2014,Ramakrishnan2015}), or optical and $\gamma$-ray light curves (e.g., \citealp{Pati2013,Cohen2014,Hovatta2014-II,Ramakrishnan2016}). An alternative approach is to organise intensive multiwavelength campaigns of individual sources (e.g., \citealp{Rani2013,Karamanavis2016}).

In this work, we revisit the optical--$\gamma$-ray \citep{Cohen2014} and $\gamma$-ray--radio correlations \citep{Max-Moerbeck2014} as well as the 
underexplored optical--radio correlation using long-term monitoring of a large sample of $\gamma$-ray-bright 
blazars. Our goal is to apply statistical analysis to reveal correlation trends and compare
them across blazar classes.  We present the sample and the results of 
basic cross-correlation analysis in Section \ref{sec:samp-cros}. In section \ref{cor_flares} we use the correlations to associate 
individual flares seen in different wavelengths and examine the statistical properties of the 
correlated and non-correlated ``orphan'' events. Section \ref{sec:disc-conc} discusses our findings 
and summarises our conclusions.

\section{Sample and Cross-Correlation Results}\label{sec:samp-cros}

\begin{figure*}
\resizebox{\hsize}{!}{\includegraphics[scale=0.5]{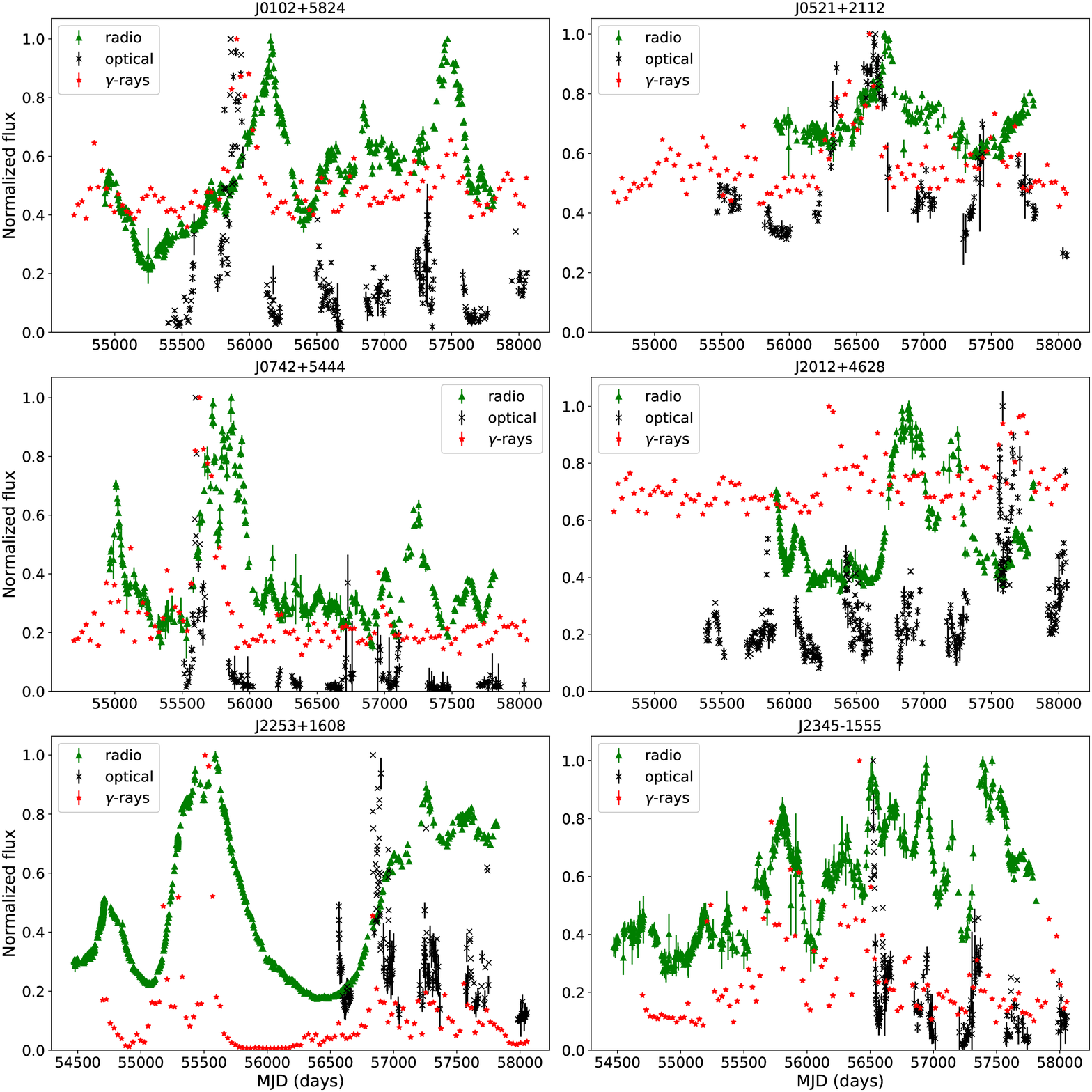} }
 \caption{Radio, optical, and $\gamma$-ray light curves of six sources with $>1.5\sigma$ correlation coefficient in all three pair combinations, normalised to the highest flux of each light curve. Green ``$\blacktriangle$'' are for radio, black ``x'' for optical, and red ``$\star$'' for $\gamma$-ray observations.}
\label{plt:example2}
\end{figure*}

We use data from the Owens Valley Radio Observatory (OVRO; 15~GHz) 40-m 
telescope\footnote{http://www.astro.caltech.edu/ovroblazars/} \citep{Richards2011}, the 0.76-m optical Katzman Automatic Imaging Telescope (KAIT; unfiltered charge-coupled device [CCD] exposures roughly corresponding to the $R$ band)
at Lick Observatory\footnote{http://herculesii.astro.berkeley.edu/kait/agn/} \citep{Filippenko2001,Li2003}, 
and the monthly averaged Large Area Telescope (LAT) $\gamma$-ray light curves from 
{\it Fermi}\footnote{https://fermi.gsfc.nasa.gov/ssc/data/access/} \citep{Acero2015}. The $\gamma$-ray light curves are automatically generated through aperture photometry using PASS8 and the latest FTOOLS package\footnote{more details on the different steps for the aperture photometry are available online, {https://fermi.gsfc.nasa.gov/ssc/data/} {analysis/scitools/{aperture\_photometry}.html}}. The photometry is in the 0.1--200~GeV range with a  $1^\circ$ radius aperture on a monthly cadence, though observations $< 5^\circ$ from the Sun have been excluded.
OVRO and KAIT have been monitoring blazars since 2007 and 2009, respectively, in support of {\it Fermi}. 
Both programs run in a fully automated mode with an approximate cadence of 3 days. The full description of the reduction pipelines for both KAIT and OVRO can be found in \cite{Li2003} and \cite{Richards2011}, respectively. The sources in our sample common to these programs
are all relatively bright objects from the first LAT blazar catalog \citep{Abdo2010-IV}.  The optical and $\gamma$-ray light curves 
are publicly available online, while the radio light curves are publicly available through the OVRO team. In this work we are considering observations from 01/2008 until 05/2017 for the radio, 07/2009 until 11/2017 for the optical, and 08/2008 until 11/2017 for the $\gamma$-rays.

Our final sample consists of the 145 common sources between OVRO, KAIT, and {\it Fermi}, 93 of which are BL Lacs, 
47 are FSRQs, and 5 are as yet unclassified sources. Sample light curves showing strong intraband correlations
are shown in Figure \ref{plt:example2}. To probe these correlations quantitatively, we calculate the  discrete correlation function 
(DCF; \citealp{Edelson1988}) for each pair of wavebands. For two time series with observations [$x_1,x_2,...x_n$], [$y_1,y_2,...y_n$],  the DCF is defined through the unbinned discrete correlation
\begin{equation}
{\rm UDCF}_{ij}=\frac{(x_i-\langle x\rangle)(y_j-\langle y\rangle)}{\sqrt{(\sigma_x^2-e_x^2)(\sigma_y^2-e_y^2)}},
\end{equation}
where ($x_i$, $y_j$) are the observations, ($\langle x\rangle$, $\langle y\rangle$) are the mean of each light curve, ($\sigma_x$, $\sigma_y$) are the standard deviation, and ($e_x$, $e_y$) are the average uncertainty, as
\begin{equation}
{\rm DCF}_\tau=\frac{1}{N}{\rm UDCF}_{ij},
\end{equation}
where $N$ is the number of ($x_i$, $y_j$) pairs for which $\tau-\Delta\tau/2\leq t_j-t_i\leq \tau+\Delta\tau$. The standard error of the DCF is defined as \citep{Edelson1988} 
\begin{equation}
\sigma_{{\rm DCF}(\tau)}=\frac{1}{N-1}\left[\sum \left({\rm UDCF}_{ij}-{\rm DCF}_\tau\right)^2  \right]^{1/2}.
\end{equation} 

We explored time lags between $-1000$ and 1000 days, a range chosen so that for any given time lag at least 2/3 of the light curves would be overlapping. This ensures a robust estimation of the DCF time lags, especially for the optical light curves where the observing gaps additionally reduce the periods of overlap with other wavelengths. The time-lag bins were set to the average cadence of the less frequently sampled light curve to ensure at least a few tens of pairs in each bin for a better determination of the uncertainty of the DCF. The entire dataset of each wavelength was used in the calculation of the DCF. Once a DCF peak 
was identified, we fitted it with a Gaussian function to best determine the cross-correlation 
coefficient at the peak as well as the peak-weighted time lag ($\tau$) and its uncertainty ($\sigma_\tau$). 

For each blazar and each waveband, we generated a distribution of random cross-correlation functions, 
using the comparison waveband curves of all other sources \citep{Cohen2014}. This generates a distribution of statistical
fluctuations in the DCF for each time-lag bin, using light curves with sampling and variability properties
similar to those of the target source. In the optical band we restrict the random cross pairs
to objects with right ascension (RA) within 3~hr. This ensures that the seasonal coverage, and hence the observation window function, are similar. This process results in $\sim5000$ false-pairs for optical--$\gamma$-ray and optical--radio cross-correlations and $\sim20,000$ false-pairs for the $\gamma$-ray--radio comparison. From this distribution of false-pair matches we estimate the median and 1$\sigma$, 2$\sigma$, and 3$\sigma$ confidence intervals for each time-lag bin. Using the false-pair distribution to measure the statistical fluctuations we quote the peak significance for each source, given as $\sigma$ confidence level (note that the ``$\sigma$'' are thresholds for the standard probability intervals; the false significant distributions are not strictly Gaussian).

\begin{figure*}
\resizebox{\hsize}{!}{\includegraphics[scale=1]{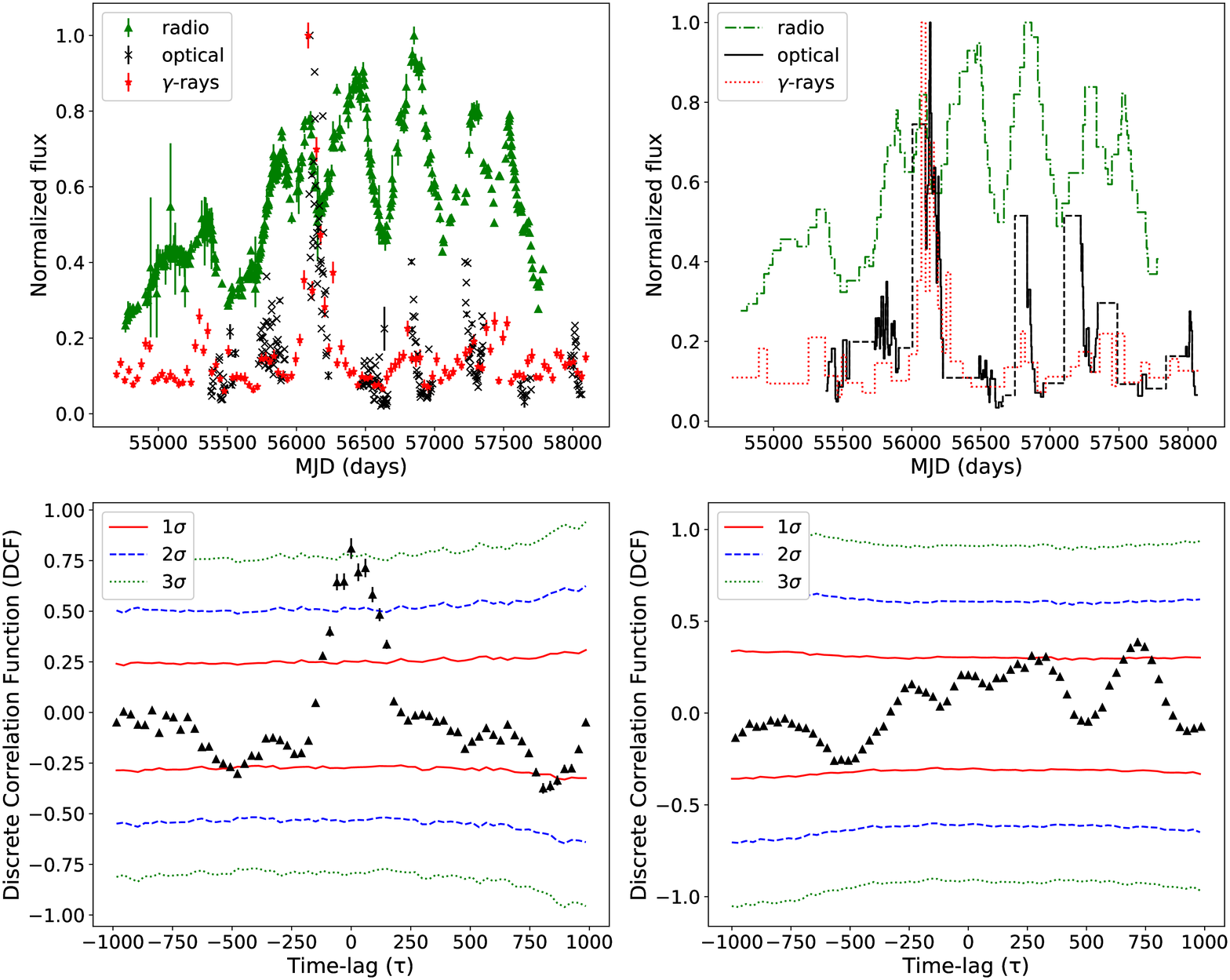} }
 \caption{{\bf Upper panel}: Radio, optical, and $\gamma$-ray light curves for J2236-1433 normalised 
to the highest flux of each light curve (left, symbols and colors as in Fig. \ref{plt:example2}) and the Bayesian block representation of the same light 
curves (right, dash-dotted green for radio, solid black for optical, dotted red for $\gamma$-rays). The dashed segments in the optical light curve denote the observing gaps. {\bf Lower panel}: Discrete correlation function for the optical--$\gamma$-ray light 
curves (left) and the $\gamma$-ray--radio light curves (right). The red solid, blue dashed, and green dotted lines mark the $1\sigma$ (68\%), $2\sigma$ (95\%), and $3\sigma$ (99.7\%) confidence intervals.}
\label{plt:example}
\end{figure*}

From the confidence intervals we estimate the significance of the peak in the real cross-correlation coefficient of interest. Figure \ref{plt:example} shows one example with the optical--$\gamma$-ray and optical--radio DCFs for J2236-1433. In this case the optical--$\gamma$-ray correlation exceeds $3\sigma$, while the
$\gamma$-ray--radio peak significance is only $\sim 1 \sigma$. The peak measurements and significances
for all pairs with a $>1\sigma$ correlation  are given in the Appendix (Table \ref{tab:results}).

\subsection{Optical--Radio Cross Correlation}

\begin{figure}
\resizebox{\hsize}{!}{\includegraphics[scale=1]{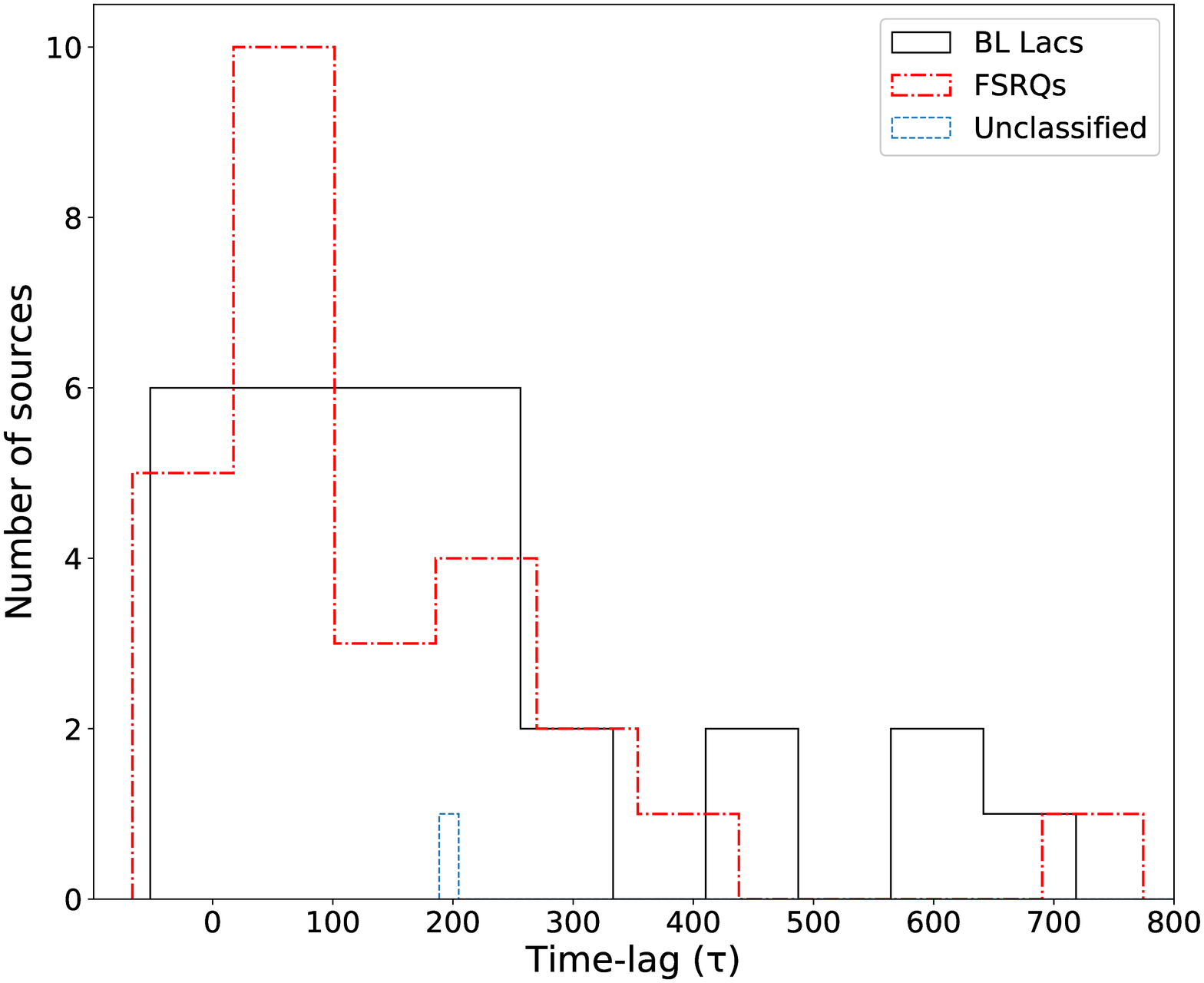} }
 \caption{Distribution of the time lags between the optical and radio emission for BL Lacs 
(solid black), FSRQs (dash-dotted red), and unclassified sources (dashed blue). Positive time lags indicate that the optical emission 
is leading the radio.}
\label{plt:lag_o-r}
\end{figure}

The optical--radio cross-correlation analysis yielded 58 sources with a $>1\sigma$ correlation 
coefficient. BL Lacs account for 31 sources, 26 are FSRQs, and 1 is unclassified. 
Figure \ref{plt:lag_o-r} shows the distribution of time lags with $>1\sigma$ correlation 
coefficients for the BL Lacs and FSRQs. 
Here positive time lags indicate optical peaks leading the radio. Clearly the majority of sources 
(89.6\%) show positive time lags. Using the two-sample Kolmogorov-Smirnov (K-S) test\footnote{The two-sample K-S test 
operates under the null hypothesis that the two samples are drawn from the same distribution. For 
any probability value $>5\%$ we cannot reject the null hypothesis.}, we do not find any significant 
difference between the BL Lac and FSRQ time lags (43.8\% probability that the two samples are drawn 
from the same distribution). There are 3 BL Lacs and 1 FSRQ with a $>3\sigma$ significant time lag. 
Out of the sources with $>3\sigma$ significant time lags all but one BL Lac (J1959+6508) have positive time lags. From the distribution of significances,
out of the 58 sources that showed at least a $>1\sigma$ correlation we estimate that 5.8 will 
be false positives; in the $>2\sigma$ set we should have no more than 0.6 false correlations.

\subsection{Optical--$\gamma$-ray Cross Correlation}
\begin{figure}
\resizebox{\hsize}{!}{\includegraphics[scale=1]{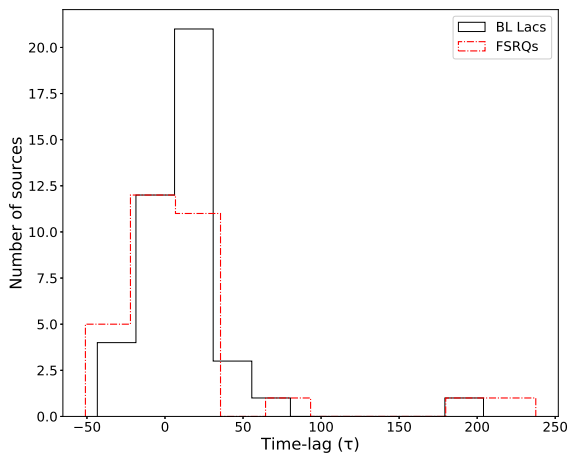} }
 \caption{Distribution of the time lags between the optical and $\gamma$-ray emission for BL Lacs 
(solid black) and FSRQs (dash-dotted red). Positive time lags indicate that the optical emission 
is leading the $\gamma$-ray.}
\label{plt:lag_o-g}
\end{figure}
\begin{figure}
\resizebox{\hsize}{!}{\includegraphics[scale=1]{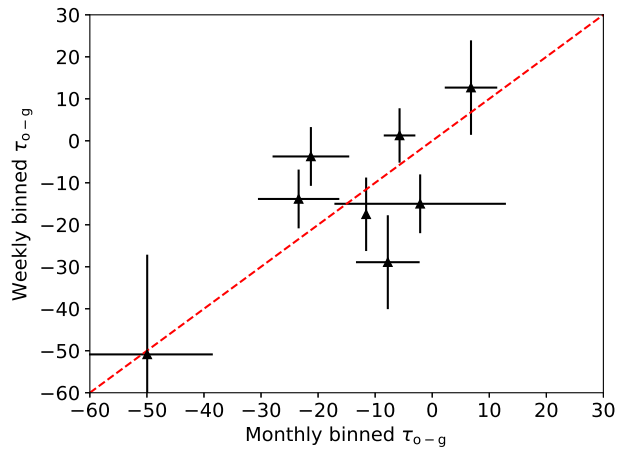} }
 \caption{Comparison between the optical--$\gamma$-ray time lags derived for 30~d and 7~d binned light curves. The red dashed line marks the one-to-one relation.}
\label{plt:30-7_comp}
\end{figure}

For the optical--$\gamma$-ray DCFs we found 73 sources (42 BL Lacs and 31 FSRQs)  with a $>1\sigma$ 
correlation peak (Fig. \ref{plt:lag_o-g}). Positive time lag indicates an optical peak leading the 
$\gamma$-rays. Again, there is no significant difference between the BL Lac and FSRQ time lags
(a K-S test yielded a 30.3\% probability). The distribution of time lags is narrow with the majority 
of sources showing time lags between [$-50$,50] days, while roughly half the sources have time lags 
consistent with zero (within 3$\sigma$), suggesting nearly coincident optical and GeV radiation zones. 
However, the 30~d cadence of the LAT light curve clearly limits
our ability to probe short timescales. In order to assess whether the 30~d sampling of the light curves is affecting the derived time lags and thus our overall results, we repeated the above analysis for eight sources with publicly available 7~d light curves included in the LAT monitored list. Figure \ref{plt:30-7_comp} shows the comparison of the derived time lags for the two different binnings of the {\it Fermi} data. The derived time lags are consistent within the uncertainties.

To investigate whether there is a systematic bias introduced by the coarser bins, we fit a line to the time lags derived by the two differently binned light curves taking the uncertainties into account. For the optical--$\gamma$-ray time lags we find a slope of $0.9\pm0.3$. For the $\gamma$-ray--radio time lags we find a slope $1.1\pm0.1$, although it should be noted that in this case there are only four sources that show a correlation (see Section \ref{rad-gam}). In both cases the intercept is consistent with zero. These tests would suggest that although it will undoubtedly be productive to repeat this analysis with a more finely sampled LAT light curve at least for the sources with bright flares, the coarser binning is not biasing our results for the time lags in any significant way.

The false positives in the optical--$\gamma$-ray case are estimated to be no more than 4.7 sources, 0.9 above $2\sigma$. There is one BL Lac object (J0045+2127) with the optical leading the $\gamma$-rays by roughly 200 days ($2\sigma$).   Considering the typical radio lag noted above, we might speculate that for this source
the $\gamma$-rays are produced closer to the radio core. However, J0045+2127 is radio-faint
and does not provide a significant $\gamma$-ray--radio DCF peak (see Table \ref{tab:results}), preventing
us from seeing the small lag that would be then expected.

\subsection{$\gamma$-ray--radio Cross Correlation}\label{rad-gam}

\begin{figure}
\resizebox{\hsize}{!}{\includegraphics[scale=1]{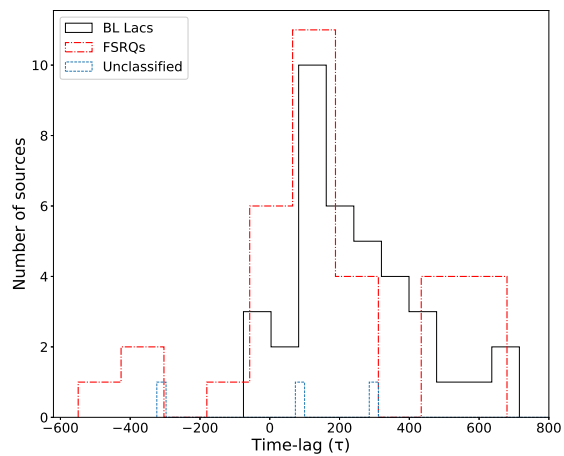} }
\caption{Distribution of the time lags between the radio and $\gamma$-ray emission for BL Lacs (solid black), FSRQs (dash-dotted red), and unclassified sources (dashed blue). Positive time lags indicate that $\gamma$-rays are leading the radio emission.}
\label{plt:lag_r-g}
\end{figure}

Finally, the $\gamma$-ray--radio cross-correlation analysis yielded 73 sources with a 
cross-correlation coefficient greater than 1$\sigma$, 37 of which are BL Lacs, 33 FSRQs, and 3 unclassified sources (Fig. \ref {plt:lag_r-g}).
Positive time lags means that the $\gamma$-rays are leading the radio emission. 
We estimate 8.2 of 73 $>1\sigma$ cross correlations and 0.5 of the $>2\sigma$ cross correlations
are false. Again, the K-S test showed no 
significant difference between the BL Lac and FSRQ time lags (15.3\% probability). The majority of sources 
(84.9\%) and all the $>3\sigma$ significance sources (2 BL Lacs) have positive time lags; 
the radio peaks are strongly delayed from the $\gamma$-ray (and optical) flares. 

By comparing the unlagged light curves of Figure \ref{plt:example2} with the correlation peak lags
in Table \ref{tab:results} of Appendix \ref{appendix}, one can visualise the expected shifts. Each of these
sources has at least $>1.5\sigma$ for all three DCF peaks.

\subsection{Comparison with Other Studies}

	Earlier investigations typically measure only a single wavelength pair, but it is worth
comparing our results with these studies' findings.

\citet{Max-Moerbeck2014} investigated the $\gamma$-ray--radio correlation using data from 
OVRO and {\it Fermi}, but with a shorter time span (about 3--4 yr). They estimated the significance of the DCF correlation coefficient by 
creating simulated light curves with statistical properties similar to those of the actual targets, using 
Monte-Carlo simulations and assuming a power-law power spectral density model. The few common sources with this work show low-significance DCF peaks ($<2\sigma$, 22 sources). There are a few additional sources for which the reported time lag is not detected in this study. Given the low significance of these correlations as well as the longer light curves considered here, we conclude that the previously reported correlations were false positive. Out of 41 sources, 
they found 3 sources with a highly significant correlation ($>2.25\sigma$) in all of which the $\gamma$-ray 
emission led the radio. Although these three sources are not included in our study, our work provides additional evidence for the $\gamma$-ray--radio lag noted there (Fig. \ref{plt:lag_r-g}).

\citet{Zhang2017} investigated the optical--radio correlation using data from KAIT 
and OVRO. They considered 70 common sources and found 55 sources with a significant correlation 
(DCF correlation coefficient $>0.5$). For the determination of the time lag and its uncertainty 
they used the flux redistribution method \citep{Peterson1998}. They also find a strong trend for
the optical to lead the radio. Of the common sources with a significant correlation, only one source (J1748+7005) appears to be inconsistent. However, it shows a low-significance DCF peak in both studies ($\sigma_{\rm DCF}=1.23$. this work; $\rm DCF=0.64$, \citealp{Zhang2017}), suggesting a higher probability for a false-positive correlation. There are a handful of sources that have a $>1\sigma$ correlation (all of which are $<2\sigma$) in our study yet do not appear to have a significant correlation in \cite[][DCF $<0.5$]{Zhang2017}, and vice versa. These discrepancies are most likely attributed to differences in the adopted methodologies. It is reassuring that most of these cases are in the low-significance range for both studies.

\citet{Cohen2014} investigated the optical--$\gamma$-ray correlation using data from 
KAIT and {\it Fermi}; our DCF methods follow this work. There are important
differences:  they only studied the 39 brightest sources in optical and $\gamma$-rays,
with a common time span of $<5$~yr. However, their use of adaptive binning \citep{Lott2012} 
allowed a better probe of small time lags $\tau$, albeit for only a few sources with very bright flares.\footnote{We should note that their sign convention for $\tau_{\rm og}$ is opposite to that used here.}
\cite{Cohen2014} find correlations for 23/39 sources at $>1\sigma$. Still, 
the finer LAT sampling from the adaptive binning may have revealed a few additional short flares, 
so such reanalysis could be productive. Of the sources in common, we generally have a substantially
higher peak DCF significance. For the common
well-detected sources, the measured lags for 19 are consistent, while the remaining sources are either listed as $<68\%$ significance or show small $\tau$ which are not well 
resolved in this work given the sampling of our LAT light curves. We do find 4 additional correlations not reported by \cite{Cohen2014}. With a larger set of sources (and more flares) 
to compare, our study is more sensitive to differences between the source classes (e.g., BL Lac/FSRQ 
differences, although no strong trends are seen even with our larger sample). 

A three-band correlation analysis was performed by \cite{Ramakrishnan2016} for 
15 sources using observations spanning 2.5~yr. They used optical, $\gamma$-rays, and two 
radio frequencies (37 and 95~GHz). Their methodology for estimating the DCF and its significance 
is similar to that of \cite{Max-Moerbeck2014}. For the two sources that show an optical--radio correlation in \cite{Ramakrishnan2016} at 37~GHz, we find larger time lags. This is expected given that 15~GHz typically probes regions farther downstream in the jet than 37~GHz. For the common sources with an optical--$\gamma$-ray correlation we find time lags consistent within the uncertainties. We have derived time lags for two sources (J0808-0751, J2232+1143) listed without a correlation in \cite{Ramakrishnan2016}, most likely owing to the longer time span of the observations considered in this work. Generally, although our radio observations are at 15~GHz, and their 
{\it Fermi} light curves are produced using weekly bins, we reach similar conclusions.

\section{Correlated Flaring Activity}\label{cor_flares}
\subsection{Flare Identification}

\begin{table}
\setlength{\tabcolsep}{11pt}
\centering
  \caption{Total and associated radio, optical, and $\gamma$-ray flares.$^a$}
  \label{tab:flares}
\begin{tabular}{@{}cccc@{}}
 \hline
   & Radio & Optical & $\gamma$-ray \\
  \hline
  Radio & 1284 & 183/433(235) & 217/686  \\
Optical & 159/903 & 2465 & 165/1361  \\
$\gamma$-ray & 159/376 & 174/384(203) & 732 \\
Median Fl. rate & 1.10 & 2.54 & 0.84\\
\hline
\end{tabular}
 $^a$Diagonal entries give the total number of flares in each waveband. Off-diagonal entries: the numerator gives the number of associated flares in the row band with flares in the column band; the denominator indicates the total number of flares in the row band; parentheses show the number of flares within periods of time with optical coverage (i.e., not within an optical observing gap). For example, in the entry in the first row, second column [183/433(235)], the numerator gives 183 associated radio flares with an optical counterpart in sources with an optical--radio correlation, the denominator indicates 433 radio flares in these sources, and the parentheses (235) give the number of flares during the optical observing season. The last row reports the median flaring rate for each wavelength in flares yr$^{-1}$ per source.
\end{table}

\begin{figure}
\resizebox{\hsize}{!}{\includegraphics[scale=1]{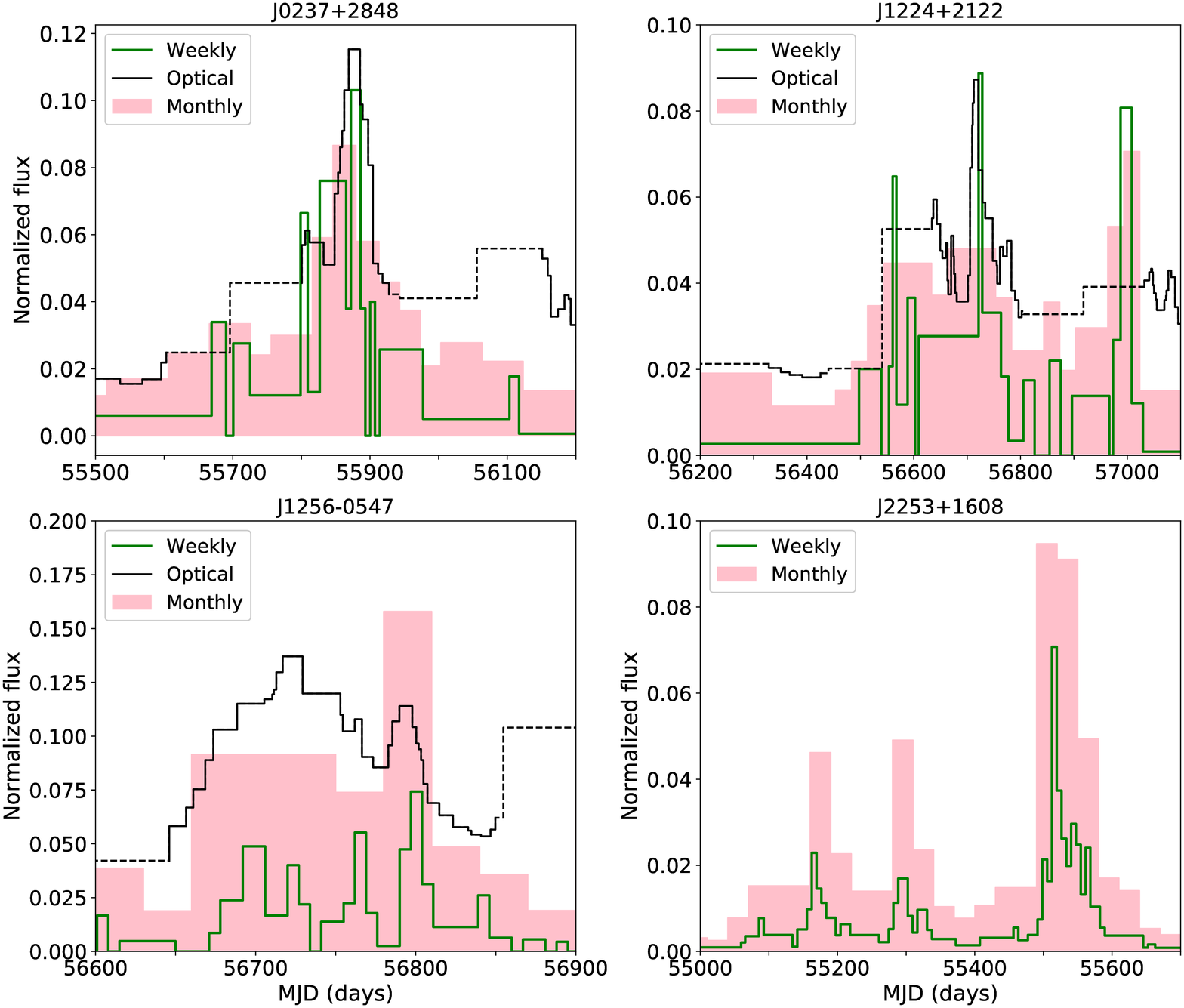} }
 \caption{Bayesian block representation of the normalised optical, 7~d, and 30~day light curves for four sources: J0237+2848 (upper left), J1224+2122 (upper right), J1256-0547 (lower left), and J2202+4216 (lower right). The light curves have been aligned according to the cross-correlation results. Solid black is for the optical, dotted green for the 7~d binned, and the red shaded area for the 30~d binned light curves. The dashed segments in the optical light curve denote the observing gaps.}
\label{plt:30-7_comp2}
\end{figure}

We also want to examine the properties and correlations of individual flaring events. To this
end we subjected the light curves to Bayesian block analysis \citep{Scargle2013}, which is a method of segmenting the light curves in ``blocks'' so that all the observations within each block are statistically consistent with a constant value. The only parameter of the method is the prior of the slope for the number of blocks, which was set to $\rm ncp\_prior\approx3$ for all three bands. Bayesian blocks allow us to model the flux variations in the light curves and obtain a relatively uniform and unbiased list of flare times and flare amplitudes in each band. Flares were identified
as local (centre of three blocks) maxima. Since every significant local maximum is counted
as a flare, this gives a large rate in highly variable bands, such as the optical.

An example Bayesian block decomposition is shown in
Figure \ref{plt:example}. This analysis yielded 1284 radio flares, 2465 optical flares, and 732 
$\gamma$-rays flares in the full survey. The large optical band flare rate may, in part, be caused by the relatively
small photometric errors adopted from KAIT aperture photometry. A few ``flares'' may also be due
to poor acquisition (bright stars in the aperture) or other systematic photometric errors. We have,
however, visually inspected the light curves to remove the most obvious such events. It is
thus interesting that there are a factor of $\sim 3.5$ more optical flares than $\gamma$-ray events.
This is even more significant since $\sim 40\%$ of the optical light curve is lost to seasonal and other
gaps, and an additional $\sim 10\%$ is close enough to gap edges to hinder flare detection. In part we expect that this is due to the coarse $\gamma$-ray sampling ($\sim30$ days vs. $\sim3$ days). 

Figure \ref{plt:30-7_comp2} shows the Bayesian block representation for four sources with available weekly binning $\gamma$-ray data. The 7~d binned light curves appear to trace the optical variability better, which could potentially improve the flare association by reducing the number of falsely matched peaks and decreasing the number of unassociated flares in the optical (see discussion below). However, overall the 30~d binned light curves adequately trace the flaring events in both the 7~d binned and optical light curves. Considering all the available 7~d light curves, we find 18\%--66\% (average $\sim 50$\%) more flares than in the 30~d light curves. Including this correction, the expected number of $\gamma$-ray flares for the entire sample is comparable to the radio, yet interestingly still less than the optical (Table \ref{tab:flares}). One can speculate that a number of low-amplitude flares have occurred in periods where only $\gamma$-ray upper limits are available from the 7~d bins; limited $\gamma$-ray counts prevent their detection.

The radio flaring rate is intermediate. Here the decrease from the optical may be attributed to the radio emission typically varying on longer timescales. Many flare complexes are seen as overlapping peaks, and so inevitably some peaks are lost in
the block analysis because of this smoothing. 

\subsubsection{Multiwavelength flaring rates}

Generally, FSRQs tend to exhibit more flares per source than the BL Lacs in radio and $\gamma$-rays, 
while BL Lacs show more flares in optical. This may be attributed to FSRQs being optically dominated by 
thermal emission from the accretion disk (e.g., \citealp{Bonning2012}). For sources with an optical--radio correlation, FSRQs 
have an average of 2.3 flares yr$^{-1}$ source$^{-1}$ in optical and 1.3 flares yr$^{-1}$ source$^{-1}$ in radio. BL Lacs 
showed 3.6 flares yr$^{-1}$ source$^{-1}$ in optical and 1.0 flares yr$^{-1}$ source$^{-1}$ in radio. FSRQs with an 
optical--$\gamma$-ray correlation displayed similar flare rates in optical (2.5 flares yr$^{-1}$ source$^{-1}$) 
and 0.8 flares yr$^{-1}$ source$^{-1}$ in $\gamma$-rays. BL Lacs exhibit a slightly higher number of flares in 
optical (3.7 flares yr$^{-1}$ source$^{-1}$) and 0.5 flares yr$^{-1}$ source$^{-1}$ in $\gamma$-rays. For the sources with 
a $\gamma$-ray--radio correlation, FSRQs show 1.1 flares yr$^{-1}$ source$^{-1}$ and 0.7 flares yr$^{-1}$ source$^{-1}$ 
and BL Lacs have 0.9 flares yr$^{-1}$ source$^{-1}$ and 0.3 flares yr$^{-1}$ source$^{-1}$ in radio and $\gamma$-rays, respectively. 
 \cite{Ramakrishnan2015} made a similar Bayesian block analysis with a smaller sample,
finding that sources with a $\gamma$-ray--radio correlation have 1.2 flares yr$^{-1}$ source$^{-1}$ in $\gamma$-rays 
and 0.6 flares yr$^{-1}$ source$^{-1}$ in radio. 

The higher flare rates seen in $\gamma$-rays for the sample 
considered by \cite{Ramakrishnan2015} may be attributed to their weekly binned {\it Fermi} light curves. Indeed, repeating the analysis on the 7~d light curves, we find that the sources show $\sim 1.3$ flares yr$^{-1}$ on average. 
Similarly, their lower flaring rate at radio wavelengths could be caused by OVRO having twice the cadence 
of the Mets\"ahovi monitoring program. If the radio flares are connected to the ejection of new jet components (e.g., \citealp{Savolainen2002}), we would expect to find a similar ejection rate from VLBI studies. Results from the MOJAVE survey for the 1.5 Jy flux-limited sample imply a lower ejection rate ($\sim0.83$ ejections yr$^{-1}$ on average; \citealp{Lister2009-2}) than suggested by the flaring rate found in this work. Either the faster cadence of the OVRO survey allows us to identify flares that are missed or are not well resolved in the VLBI maps, or a small fraction of these flares are random fluctuations not connected to the ejection of a new jet component. On the other hand, at 43~GHz the ejection rate is higher ($\sim1.25$ ejections yr$^{-1}$ on average; \citealp{Jorstad2017}). Given the longer radio time response at 15~GHz, this suggests that some individual events identified in this work might in fact be the superposition of multiple flares that could in principle be resolved at a higher frequency.

Sources that have a significant correlation between at least one pair of wavelengths show on 
average 1.1 flares yr$^{-1}$ source$^{-1}$ in radio, 2.4 flares yr$^{-1}$ source$^{-1}$ in optical, and 0.5 flares yr$^{-1}$ source$^{-1}$ 
in $\gamma$-rays. Sources that did not exhibit any significant multiwavelength correlation have 
a lower flare rate on average (0.6 flares yr$^{-1}$ source$^{-1}$ in radio, 2.2 flares yr$^{-1}$ source$^{-1}$ in optical, and 0.2 flares yr$^{-1}$ source$^{-1}$ in $\gamma$-rays). Of course, this may be a selection effect, as in sources that show fewer flaring events the light-curve fluctuations are more likely to be dominated by uncorrelated stochastic variations weakening the DCF peaks. Overall, there are 5.4\% less optical 
flares, 42\% less radio, and 59\% less $\gamma$-ray events in sources without any significant correlation compared to sources with a significant correlation between at least one pair of wavelengths.

\subsubsection{Associated and Orphan flares}

	After identifying the flares, we focus on the sources showing at least one significant
interband correlation. This allows us to align the light curves using the measured DCF $\tau$
and make interband identifications. We consider a flare to be coincident if the Bayesian block
of the peak bins overlaps in the two wavebands.
We are also interested in ``orphan'' flares, seen in one band but not another. These have been
noted in past studies (e.g., \citealp{Krawczynski2004,Rani2013}), but it has not previously been possible to make statistical statements
about the fraction of unassociated events.
Table \ref{tab:flares} gives the number of coincident flares in each waveband over the number of
flares in correlated sources for the comparison waveband. For the optical we must exclude the
time in the seasonal gaps, so we list in parenthesis the number of flares not in gaps.

	The first statistic to note is that the high optical flare rate leads to a high coincidence
rate -- fully 78\% of the radio flares in correlated optical sources during the observing window
have correlated optical flares. Similarly 86\% of $\gamma$-ray flares during correlated active
optical observations have an optical counterpart. With the wide radio blocks and the 30~d $\gamma$-ray
sampling, some of these associations are doubtless false coincidences. 
 To quantify this false-pair pollution, we repeat the same analysis for sources with at 
least a 2$\sigma$ significant correlation but randomly misalign the light curves with shifts of up to
300~d and compute the number of ``associated'' flares. This suggests that roughly 18\% of radio and 28\% 
of $\gamma$-ray flares could be falsely associated with an optical counterpart. The higher percentage 
of false $\gamma$-ray associations is most likely caused by the lower cadence and coarser blocking of 
the $\gamma$-ray light curves. For sources with a $\gamma$-ray--radio correlation, fewer than 
10\% of either radio or $\gamma$-ray flares could be falsely associated. Still, we infer that we have identified true
associations for $\sim 60$\% of the radio and $\gamma$-ray flares. Unsurprisingly, with
the high optical flare rate most are unassociated, with only 18\% showing radio counterparts
and 12\% $\gamma$-ray counterparts.  In contrast, comparing the radio and $\gamma$-rays directly, we see
that 32\% of the radio flares show associated $\gamma$-ray events, while 42\% of the 
$\gamma$-ray flares appear also in the radio. We conclude that (at least to our nonuniform 
sensitivity) over half of the radio/$\gamma$-ray events are mutual orphans. While most events
have optical associations, and while the fact that we get highly significant correlations means
that a fair fraction of these must be real, the high density of optical events does introduce some false pairs. 

BL Lacs and FSRQs show a similar percentage of associated radio ($\sim77\%$) and $\gamma$-ray (83--88\%) 
flares with an optical counterpart. FSRQs have roughly 20\% of their optical flares associated with 
either a radio or  a $\gamma$-ray flare. The percentages are significantly lower for BL Lacs 
(15.9\% associations with radio and 8.4\% associations with $\gamma$-ray flares). For sources 
with a $\gamma$-ray--radio correlation, 36--41\% of flares in FSRQs are associated between the 
two wavelengths. BL Lacs have fewer associated radio (24.7\%) than $\gamma$-ray flares (42.8\%).

Repeating the flare-association analysis on the 7~d light curves, we find an increase in the orphan $\gamma$-ray events for the sources with a $\gamma$-ray--radio correlation (although in this case only four sources are considered). For the sources with an optical--$\gamma$-ray correlation there is an improvement in the flare association. We find that roughly 24\% of the optical and 91\% of the $\gamma$-ray flares (an improvement of $\sim$+6\%) are now associated with a counterpart. It is interesting that even with a factor of 4 improvement in the sampling of the $\gamma$-ray light curves, there is only a small reduction in the number of orphan optical events. This suggests that although repeating the analysis with more finely binned light curves for the entire sample would undoubtedly be beneficial, our statistical results and conclusions are not strongly affected by the choice of coarser bins.

\subsection{Flare Amplitude Distributions}

	Our Bayesian block analysis also lets us compare the flare amplitude distributions. We compute
the flare flux density by computing the weighted average flux density within the lowest 20\% of the entire light 
curve as a background/quiescence level and subtracting this from the flux density in the peak block. We can then compare the amplitude distributions of various sources (and source groups) using
the Wilcoxon rank-sum (WRS) test\footnote{The WRS test is similar to the K-S test with the alternate 
hypothesis that one sample has systematically higher values than the other.}.
	
One useful comparison is between the associated and orphan flares. For example, taking
optical--radio and optical--$\gamma$-ray correlated sources, the associated flares are 
distinctly brighter than the orphan sets (WRS $p$-value $<10^{-4}\%$ for all these bands). In contrast, for the $\gamma$-ray--radio correlated 
sources the orphan $\gamma$-ray flares are consistent with the correlated set (30.9\%) while 
the associated radio flares are brighter at a statistically significant ($P\approx10^{-13}$\%) level. This would hint that the mechanisms driving the two wavebands' events are not strictly proportional.

\section{Discussion and Conclusions}\label{sec:disc-conc}

In this work we have investigated the temporal correlations between the optical, radio, and $\gamma$-ray 
emission of a large sample of blazars. Out of the 145 sources, 105 revealed at least one $>1\sigma$ 
significant correlation coefficient between wavebands. Out of these, 38 showed a correlation 
in only one pair of wavelengths, 35 in two pairs, and 32 in all three pairs. Based 
on the significance of DCF peaks, we estimated that only 9.8\% of the optical--radio, 
6.4\% of the optical--$\gamma$-ray, and 11.2\% of the $\gamma$-ray--radio cross correlations 
will be false positives. For the sources with a $>2\sigma$ significant correlation the 
false-positive rate is roughly 2\% in any wavelength pair. It should be noted that blazars exhibit a wide variety of flaring patterns and behaviours which could be the result of different mechanisms operating even in the same source (e.g., \citealp{Chen2012,Liodakis2017}). It is possible for a source to show both correlated and uncorrelated events with other wavelengths, which can impact the significance of DCF. This could explain why many of our sources show only a 1$\sigma$ or 2$\sigma$ significant DCF peaks. Our results are qualitatively consistent with those of previous studies, but our longer time base has allowed us to identify a larger number of, and more significant, correlations. Continued monitoring would doubtless improve the situation further.

We find that the optical--$\gamma$-ray time lags are generally small, while both lead the radio 
by $\sim30$--150~d. However, there are three sources (J0433+2905, J1849+6705, and J2236+2828) in which 
all three wavelengths are correlated with lags $<30$~d. In fact, unusually, J0433+2905 shows the 
optical emission coming 20--30~d earlier than both radio and $\gamma$-rays (the other two are 
more conventional). This suggests that the emission regions are located in close proximity. The fact that there is a strong connection between optical and $\gamma$-ray variations favours leptonic (i.e., inverse-Compton scattering of the optical photons as the production mechanism for the $\gamma$-ray emission) over hadronic processes. On the other hand, the fact that the radio usually lags all other wavelengths suggests that it is typically downstream from both the optical and $\gamma$-ray emission regions. Since the radio variations are connected to the ejection of new components from the radio core seen in VLBI maps \citep{Savolainen2002}, this would place the $\gamma$-ray emission regions between the supermassive black hole and the radio core. Combined with the generally longer timescale variations seen at radio wavelengths (e.g., \citealp{Hovatta2007}), our results favour emission scenarios of an expanding disturbance propagating in the jet and becoming optically thin at higher frequencies before becoming transparent at radio wavelengths (e.g., \citealp{Marscher1985,Max-Moerbeck2014}).
 
Using Bayesian block decomposition \citep{Scargle2013}, we have studied a large number
of individual flares and their multiwavelength properties. We have relied on the DCF analysis to
align these light curves, allowing cross-band identification of individual flaring events. This has
also for the first time allowed a robust determination of orphan (to our sensitivity) flares in
all three wavebands.

	Overall, sources showing a lower flaring rate tend to have less significant interband correlations,
but this may be a simple selection effect. Comparing BL Lacs and FSRQs, the former show a higher
flaring rate in the optical (3.6--3.7 vs. 2.3--2.5 flares yr$^{-1}$ source$^{-1}$ on average), while showing lower
radio (0.9--1 vs. 1.1--1.3 flares yr$^{-1}$ source$^{-1}$) and $\gamma$-ray (0.3--0.5 vs. 0.6--0.8 flares yr$^{-1}$ source$^{-1}$) activity.

Two main effects limit the present analysis. First, seasonal gaps in the optical light curves limit
the number of flare identifications and make long-term trends difficult to follow. Although observational gaps in the optical are unavoidable because of the Sun, in many cases these are enlarged (or even induced) by weather and/or technical related constraints. Multisite monitoring could help minimise the extent of said gaps. Second, we have used only publicly accessible $\gamma$-ray light curves, with coarse 30~d sampling. As shown in Figures \ref{plt:30-7_comp} and \ref{plt:30-7_comp2} (see also the discussion in Section \ref{cor_flares}), our overall statistical results and conclusions should not be strongly affected by the choice of bin size; however, finer sampling such as the adaptive binning used by \citet{Cohen2014} can find more flares and probe shorter timescales . For this reason, our results on the flare rates and flare associations with respect to the $\gamma$-ray light curves should be treated as limits. We are currently pursuing a more detailed analysis of the $\gamma$-ray light curves that will allow us to probe time delays and flare associations on shorter timescales between optical and $\gamma$-rays. 

In summary, our results show the following.

\begin{itemize}

\item {The radio emission generally lags the optical/$\gamma$-rays, suggesting that the higher
energy radiation arises inward of the radio cores of the jets. }

\item {The optical emission is closely connected to the $\gamma$-ray emission, with roughly half the 
sources having time lags consistent with zero.}

\item {A few sources seem to have all three bands colocated (e.g., J0433+2905,  J2236+2828).}
 
\item {Low radio and $\gamma$-ray activity likely explains the lack of significant correlation for many of
our sources.}

\item We found no significant difference between associated and orphan $\gamma$-ray flares in sources 
with a significant $\gamma$-ray--radio correlation. In all other cases (and wavelengths), flares 
have higher flux when associated with the other band than when they are orphans.

\end{itemize}

\section*{Acknowledgments}

We thank the anonymous referee and T. Hovatta for comments and suggestions that helped improve this work. The \textit{Fermi} LAT Collaboration acknowledges generous ongoing support from a number of agencies and institutes that have supported both the development and the operation of the LAT as well as scientific data analysis.  These include the National Aeronautics and Space Administration (NASA) and the Department of Energy in the United States, the Commissariat \`a l'Energie Atomique and the Centre National de la Recherche Scientifique/Institut National de Physique Nucl\'eaire et de Physique des Particules in France, the Agenzia Spaziale Italiana and the Istituto Nazionale di Fisica Nucleare in Italy, the Ministry of Education, Culture, Sports, Science, and Technology (MEXT), High Energy Accelerator Research Organization (KEK), and Japan Aerospace Exploration Agency (JAXA) in Japan, and the K.~A.~Wallenberg Foundation, the Swedish Research Council, and the Swedish National Space Board in Sweden. Additional support for science analysis during the operations phase is gratefully acknowledged from the Istituto Nazionale di Astrofisica in Italy and the Centre National d'\'Etudes Spatiales in France.

This research has made use of data from the robotic 0.76-m Katzman Automatic Imaging telescope \citep{Li2003} at Lick Observatory. We thank the late Weidong Li for setting up the KAIT blazar monitoring program. This work was financed in part by NASA grants NNX10AU09G, GO-31089, NNX12AF12G, and NAS5-00147. A.V.F. and W.Z. are also grateful for support from the Christopher R. Redlich Fund, the TABASGO Foundation, NSF grant AST-1211916, and the Miller Institute for Basic Research in Science (UC Berkeley).  KAIT and its ongoing operation were made possible by donations from Sun Microsystems, Inc., the Hewlett-Packard Company, AutoScope Corporation, Lick Observatory, the NSF, the University of California, the Sylvia and Jim Katzman Foundation, and the TABASGO Foundation. Research at Lick Observatory is partially supported by a generous gift from Google. This work has made use of data from the OVRO 40-m monitoring program \citep{Richards2011}, which is supported in part by NASA grants NNX08AW31G, NNX11A043G, and NNX14AQ89G, as well as by NSF grants AST-0808050 and AST-1109911. A.V.F.'s research was conducted in part at the Aspen Center for Physics, which is supported by NSF grant PHY-1607611; he thanks the Center for its hospitality during the supermassive black holes workshop in June and July 2018. 

\bibliographystyle{mnras}
\bibliography{bibliography} 

\begin{thebibliography}{}
\makeatletter
\relax
\def\mn@urlcharsother{\let\do\@makeother \do\$\do\&\do\#\do\^\do\_\do\%\do\~}
\def\mn@doi{\begingroup\mn@urlcharsother \@ifnextchar [ {\mn@doi@}
  {\mn@doi@[]}}
\def\mn@doi@[#1]#2{\def\@tempa{#1}\ifx\@tempa\@empty \href
  {http://dx.doi.org/#2} {doi:#2}\else \href {http://dx.doi.org/#2} {#1}\fi
  \endgroup}
\def\mn@eprint#1#2{\mn@eprint@#1:#2::\@nil}
\def\mn@eprint@arXiv#1{\href {http://arxiv.org/abs/#1} {{\tt arXiv:#1}}}
\def\mn@eprint@dblp#1{\href {http://dblp.uni-trier.de/rec/bibtex/#1.xml}
  {dblp:#1}}
\def\mn@eprint@#1:#2:#3:#4\@nil{\def\@tempa {#1}\def\@tempb {#2}\def\@tempc
  {#3}\ifx \@tempc \@empty \let \@tempc \@tempb \let \@tempb \@tempa \fi \ifx
  \@tempb \@empty \def\@tempb {arXiv}\fi \@ifundefined
  {mn@eprint@\@tempb}{\@tempb:\@tempc}{\expandafter \expandafter \csname
  mn@eprint@\@tempb\endcsname \expandafter{\@tempc}}}

\bibitem[\protect\citeauthoryear{{Abdo} et~al.,}{{Abdo}
  et~al.}{2010a}]{Abdo2010-IV}
{Abdo} A.~A.,  et~al., 2010a, \mn@doi [\apj] {10.1088/0004-637X/715/1/429},
  \href {http://adsabs.harvard.edu/abs/2010ApJ...715..429A} {715, 429}

\bibitem[\protect\citeauthoryear{{Abdo} et~al.,}{{Abdo}
  et~al.}{2010b}]{Abdo2010-II}
{Abdo} A.~A.,  et~al., 2010b, \mn@doi [\apj] {10.1088/0004-637X/716/1/30},
  \href {http://adsabs.harvard.edu/abs/2010ApJ...716...30A} {716, 30}

\bibitem[\protect\citeauthoryear{{Acero} et~al.,}{{Acero}
  et~al.}{2015}]{Acero2015}
{Acero} F.,  et~al., 2015, \mn@doi [\apjs] {10.1088/0067-0049/218/2/23}, \href
  {http://adsabs.harvard.edu/abs/2015ApJS..218...23A} {218, 23}

\bibitem[\protect\citeauthoryear{{Blinov} et~al.,}{{Blinov}
  et~al.}{2017}]{Blinov2017}
{Blinov} D.,  et~al., 2017, preprint, \href
  {http://adsabs.harvard.edu/abs/2017arXiv171008922B} {} (\mn@eprint {arXiv}
  {1710.08922})

\bibitem[\protect\citeauthoryear{{Bonning} et~al.,}{{Bonning}
  et~al.}{2012}]{Bonning2012}
{Bonning} E.,  et~al., 2012, \mn@doi [\apj] {10.1088/0004-637X/756/1/13}, \href
  {http://adsabs.harvard.edu/abs/2012ApJ...756...13B} {756, 13}

\bibitem[\protect\citeauthoryear{{Chen}, {Fossati}, {B{\"o}ttcher}  \&
  {Liang}}{{Chen} et~al.}{2012}]{Chen2012}
{Chen} X.,  {Fossati} G.,  {B{\"o}ttcher} M.,   {Liang} E.,  2012, \mn@doi
  [\mnras] {10.1111/j.1365-2966.2012.21283.x}, \href
  {http://adsabs.harvard.edu/abs/2012MNRAS.424..789C} {424, 789}

\bibitem[\protect\citeauthoryear{{Cohen}, {Romani}, {Filippenko}, {Cenko},
  {Lott}, {Zheng}  \& {Li}}{{Cohen} et~al.}{2014}]{Cohen2014}
{Cohen} D.~P.,  {Romani} R.~W.,  {Filippenko} A.~V.,  {Cenko} S.~B.,  {Lott}
  B.,  {Zheng} W.,   {Li} W.,  2014, \mn@doi [\apj]
  {10.1088/0004-637X/797/2/137}, \href
  {http://adsabs.harvard.edu/abs/2014ApJ...797..137C} {797, 137}

\bibitem[\protect\citeauthoryear{{Dermer}, {Schlickeiser}  \&
  {Mastichiadis}}{{Dermer} et~al.}{1992}]{Dermer1992}
{Dermer} C.~D.,  {Schlickeiser} R.,   {Mastichiadis} A.,  1992, \aap, \href
  {http://adsabs.harvard.edu/abs/1992A%26A...256L..27D} {256, L27}

\bibitem[\protect\citeauthoryear{{Edelson} \& {Krolik}}{{Edelson} \&
  {Krolik}}{1988}]{Edelson1988}
{Edelson} R.~A.,  {Krolik} J.~H.,  1988, \mn@doi [\apj] {10.1086/166773}, \href
  {http://adsabs.harvard.edu/abs/1988ApJ...333..646E} {333, 646}

\bibitem[\protect\citeauthoryear{{Filippenko}, {Li}, {Treffers}  \&
  {Modjaz}}{{Filippenko} et~al.}{2001}]{Filippenko2001}
{Filippenko} A.~V.,  {Li} W.~D.,  {Treffers} R.~R.,   {Modjaz} M.,  2001, in
  {Paczynski} B.,  {Chen} W.-P.,   {Lemme} C.,  eds,  Astronomical Society of
  the Pacific Conference Series Vol. 246, IAU Colloq. 183: Small Telescope
  Astronomy on Global Scales. p.~121

\bibitem[\protect\citeauthoryear{{Fuhrmann} et~al.,}{{Fuhrmann}
  et~al.}{2014}]{Fuhrmann2014}
{Fuhrmann} L.,  et~al., 2014, \mn@doi [\mnras] {10.1093/mnras/stu540}, \href
  {http://adsabs.harvard.edu/abs/2014MNRAS.441.1899F} {441, 1899}

\bibitem[\protect\citeauthoryear{{Hovatta}, {Tornikoski}, {Lainela}, {Lehto},
  {Valtaoja}, {Torniainen}, {Aller}  \& {Aller}}{{Hovatta}
  et~al.}{2007}]{Hovatta2007}
{Hovatta} T.,  {Tornikoski} M.,  {Lainela} M.,  {Lehto} H.~J.,  {Valtaoja} E.,
  {Torniainen} I.,  {Aller} M.~F.,   {Aller} H.~D.,  2007, \mn@doi [A\&A]
  {10.1051/0004-6361:20077529}, \href
  {http://adsabs.harvard.edu/abs/2007A%26A...469..899H} {469, 899}

\bibitem[\protect\citeauthoryear{{Hovatta} et~al.,}{{Hovatta}
  et~al.}{2014}]{Hovatta2014-II}
{Hovatta} T.,  et~al., 2014, \mn@doi [\mnras] {10.1093/mnras/stt2494}, \href
  {http://adsabs.harvard.edu/abs/2014MNRAS.439..690H} {439, 690}

\bibitem[\protect\citeauthoryear{{Jorstad} et~al.,}{{Jorstad}
  et~al.}{2017}]{Jorstad2017}
{Jorstad} S.~G.,  et~al., 2017, \mn@doi [\apj] {10.3847/1538-4357/aa8407},
  \href {http://adsabs.harvard.edu/abs/2017ApJ...846...98J} {846, 98}

\bibitem[\protect\citeauthoryear{{Karamanavis} et~al.,}{{Karamanavis}
  et~al.}{2016}]{Karamanavis2016}
{Karamanavis} V.,  et~al., 2016, \mn@doi [\aap] {10.1051/0004-6361/201527796},
  \href {http://adsabs.harvard.edu/abs/2016A%26A...590A..48K} {590, A48}

\bibitem[\protect\citeauthoryear{{Krawczynski} et~al.,}{{Krawczynski}
  et~al.}{2004}]{Krawczynski2004}
{Krawczynski} H.,  et~al., 2004, \mn@doi [\apj] {10.1086/380393}, \href
  {http://adsabs.harvard.edu/abs/2004ApJ...601..151K} {601, 151}

\bibitem[\protect\citeauthoryear{{Li}, {Filippenko}, {Chornock}  \& {Jha}}{{Li}
  et~al.}{2003}]{Li2003}
{Li} W.,  {Filippenko} A.~V.,  {Chornock} R.,   {Jha} S.,  2003, \mn@doi
  [\pasp] {10.1086/376432}, \href
  {http://adsabs.harvard.edu/abs/2003PASP..115..844L} {115, 844}

\bibitem[\protect\citeauthoryear{{Liodakis} et~al.,}{{Liodakis}
  et~al.}{2017}]{Liodakis2017}
{Liodakis} I.,  et~al., 2017, \mn@doi [\mnras] {10.1093/mnras/stx002}, \href
  {http://adsabs.harvard.edu/abs/2017MNRAS.466.4625L} {466, 4625}

\bibitem[\protect\citeauthoryear{{Lister} et~al.,}{{Lister}
  et~al.}{2009}]{Lister2009-2}
{Lister} M.~L.,  et~al., 2009, \mn@doi [\aj] {10.1088/0004-6256/138/6/1874},
  \href {http://adsabs.harvard.edu/abs/2009AJ....138.1874L} {138, 1874}

\bibitem[\protect\citeauthoryear{{Lott}, {Escande}, {Larsson}  \&
  {Ballet}}{{Lott} et~al.}{2012}]{Lott2012}
{Lott} B.,  {Escande} L.,  {Larsson} S.,   {Ballet} J.,  2012, \mn@doi [\aap]
  {10.1051/0004-6361/201218873}, \href
  {http://adsabs.harvard.edu/abs/2012A%26A...544A...6L} {544, A6}

\bibitem[\protect\citeauthoryear{{Marscher} \& {Gear}}{{Marscher} \&
  {Gear}}{1985}]{Marscher1985}
{Marscher} A.~P.,  {Gear} W.~K.,  1985, \mn@doi [\apj] {10.1086/163592}, \href
  {http://adsabs.harvard.edu/abs/1985ApJ...298..114M} {298, 114}

\bibitem[\protect\citeauthoryear{{Marscher} et~al.,}{{Marscher}
  et~al.}{2008}]{Marscher2008}
{Marscher} A.~P.,  et~al., 2008, \mn@doi [\nat] {10.1038/nature06895}, \href
  {http://adsabs.harvard.edu/abs/2008Natur.452..966M} {452, 966}

\bibitem[\protect\citeauthoryear{{Max-Moerbeck} et~al.,}{{Max-Moerbeck}
  et~al.}{2014}]{Max-Moerbeck2014}
{Max-Moerbeck} W.,  et~al., 2014, \mn@doi [\mnras] {10.1093/mnras/stu1749},
  \href {http://adsabs.harvard.edu/abs/2014MNRAS.445..428M} {445, 428}

\bibitem[\protect\citeauthoryear{{Pati{\~n}o-{\'A}lvarez}, {Carrami{\~n}ana},
  {Carrasco}  \& {Chavushyan}}{{Pati{\~n}o-{\'A}lvarez}
  et~al.}{2013}]{Pati2013}
{Pati{\~n}o-{\'A}lvarez} V.,  {Carrami{\~n}ana} A.,  {Carrasco} L.,
  {Chavushyan} V.,  2013, preprint, \href
  {http://adsabs.harvard.edu/abs/2013arXiv1303.1898P} {} (\mn@eprint {arXiv}
  {1303.1898})

\bibitem[\protect\citeauthoryear{{Peterson}, {Wanders}, {Horne}, {Collier},
  {Alexander}, {Kaspi}  \& {Maoz}}{{Peterson} et~al.}{1998}]{Peterson1998}
{Peterson} B.~M.,  {Wanders} I.,  {Horne} K.,  {Collier} S.,  {Alexander} T.,
  {Kaspi} S.,   {Maoz} D.,  1998, \mn@doi [\pasp] {10.1086/316177}, \href
  {http://adsabs.harvard.edu/abs/1998PASP..110..660P} {110, 660}

\bibitem[\protect\citeauthoryear{{Ramakrishnan}, {Hovatta}, {Nieppola},
  {Tornikoski}, {L{\"a}hteenm{\"a}ki}  \& {Valtaoja}}{{Ramakrishnan}
  et~al.}{2015}]{Ramakrishnan2015}
{Ramakrishnan} V.,  {Hovatta} T.,  {Nieppola} E.,  {Tornikoski} M.,
  {L{\"a}hteenm{\"a}ki} A.,   {Valtaoja} E.,  2015, \mn@doi [\mnras]
  {10.1093/mnras/stv321}, \href
  {http://adsabs.harvard.edu/abs/2015MNRAS.452.1280R} {452, 1280}

\bibitem[\protect\citeauthoryear{{Ramakrishnan} et~al.,}{{Ramakrishnan}
  et~al.}{2016}]{Ramakrishnan2016}
{Ramakrishnan} V.,  et~al., 2016, \mn@doi [\mnras] {10.1093/mnras/stv2653},
  \href {http://adsabs.harvard.edu/abs/2016MNRAS.456..171R} {456, 171}

\bibitem[\protect\citeauthoryear{{Rani} et~al.,}{{Rani}
  et~al.}{2013}]{Rani2013}
{Rani} B.,  et~al., 2013, \mn@doi [\aap] {10.1051/0004-6361/201321058}, \href
  {http://adsabs.harvard.edu/abs/2013A%26A...552A..11R} {552, A11}

\bibitem[\protect\citeauthoryear{{Richards} et~al.,}{{Richards}
  et~al.}{2011}]{Richards2011}
{Richards} J.~L.,  et~al., 2011, \mn@doi [\apjs] {10.1088/0067-0049/194/2/29},
  \href {http://adsabs.harvard.edu/abs/2011ApJS..194...29R} {194, 29}

\bibitem[\protect\citeauthoryear{{Savolainen}, {Wiik}, {Valtaoja}, {Jorstad}
  \& {Marscher}}{{Savolainen} et~al.}{2002}]{Savolainen2002}
{Savolainen} T.,  {Wiik} K.,  {Valtaoja} E.,  {Jorstad} S.~G.,   {Marscher}
  A.~P.,  2002, \mn@doi [\aap] {10.1051/0004-6361:20021236}, \href
  {http://adsabs.harvard.edu/abs/2002A%26A...394..851S} {394, 851}

\bibitem[\protect\citeauthoryear{{Scargle}, {Norris}, {Jackson}  \&
  {Chiang}}{{Scargle} et~al.}{2013}]{Scargle2013}
{Scargle} J.~D.,  {Norris} J.~P.,  {Jackson} B.,   {Chiang} J.,  2013, \mn@doi
  [\apj] {10.1088/0004-637X/764/2/167}, \href
  {http://adsabs.harvard.edu/abs/2013ApJ...764..167S} {764, 167}

\bibitem[\protect\citeauthoryear{Zhang, Zhao, Zhang  \& Dai}{Zhang
  et~al.}{2017}]{Zhang2017}
Zhang B.~K.,  Zhao X.~Y.,  Zhang L.,   Dai B.~Z.,  2017, The Astrophysical
  Journal Supplement Series, 231, 14

\makeatother
\end{thebibliography}

\appendix
\section{Cross-Correlation Results}\label{appendix}

\begin{table*}
\setlength{\tabcolsep}{8pt}
\centering
  \caption{Cross-correlation results for the sources in our sample that showed a $>1\sigma$ significant DCF peak. Columns: (1) KAIT name, (2) class (B for BL Lacs, F for FSRQs), (3) redshift, (4) optical--radio time lag ($\tau_{\rm o-r}$), (5) uncertainty of $\tau_{\rm o-r}$, (6) significance of the peak $\rm DCF_{\rm o-r}$ coefficient, (7) optical--$\gamma$-rays time lag ($\tau_{\rm o-g}$), (8) uncertainty of $\tau_{\rm o-g}$, (9) significance of the peak $\rm DCF_{\rm o-g}$ coefficient, (10) $\gamma$-ray--radio time lag ($\tau_{\rm g-r}$), (11) uncertainty of $\tau_{\rm g-r}$, (12) significance of the peak $\rm DCF_{\rm g-r}$ coefficient. For a positive $\tau_{\rm o-r}$ or $\tau_{\rm o-g}$, the optical emission is leading the radio or $\gamma$-rays, respectively; for a positive  $\tau_{\rm g-r}$ the $\gamma$-ray emission is leading the radio.}
  \label{tab:results}
\begin{tabular}{@{}llcccccccccc@{}}
\hline
Name & Class & z & $\tau_{\rm o-r}$ & $\sigma\tau_{\rm o-r}$ & Signif.  & $\tau_{\rm o-g}$ & $\sigma\tau_{\rm o-g}$ & Signif. &$\tau_{\rm g-r}$ & $\sigma\tau_{\rm g-r}$ & Signif.\\
& & & & & (${\rm DCF_{\rm o-r}}$) & & & (${\rm DCF_{\rm o-g}}$) & & & (${\rm DCF_{\rm g-r}}$)\\
  (1) & (2) & (3) & (4) & (5) & (6) & (7) & (8) & (9) & (10) & (11) & (12) \\
  \hline
J0017-0512 & F & 0.227 & 200.96 & 1.2 & 2.04 & 20.23 & 3.66 & 2.1 & 113.15 & 5.96 & 1.06 \\ 
J0033-1921 & B & 0.61 & - & - & - & - & - & - & - & - & - \\ 
J0035+1515 & B & 1.409 & - & - & - & - & - & - & 365.7 & 5.96 & 1.9 \\ 
J0035+5950 & B & 0.086 & - & - & - & 15.11 & 0.0 & 1.01 & 221.9 & 29.41 & 2.43 \\ 
J0045+2127 & B & - & - & - & - & 204.06 & 2.65 & 2.0 & - & - & - \\ 
J0050-0929 & B & 0.635 & - & - & - & -17.88 & 3.03 & 3.05 & 161.31 & 2.34 & 2.8 \\ 
J0102+5824 & F & 0.644 & 270.71 & 2.26 & 2.2 & -2.16 & 9.2 & 3.22 & 162.41 & 5.13 & 1.96 \\ 
J0112+2244 & B & 0.265 & - & - & - & 7.11 & 4.0 & 2.53 & - & - & - \\ 
J0115+2519 & B & 0.37 & - & - & - & 13.59 & 12.97 & 2.08 & - & - & - \\ 
J0132-1654 & F & 1.02 & - & - & - & -35.28 & 5.74 & 2.02 & -12.76 & 9.7 & 1.13 \\ 
J0136+3905 & B & 0.75 & - & - & - & - & - & - & - & - & - \\ 
J0141-0928 & B & 0.733 & -18.49 & 4.22 & 1.28 & 24.61 & 0.88 & 2.82 & - & - & - \\ 
J0154+0823 & B & 0.681 & 178.02 & 0.22 & 1.0 & - & - & - & 343.01 & 12.08 & 2.47 \\ 
J0204-1701 & F & 1.74 & 774.39 & 3.89 & 2.41 & 6.26 & 2.6 & 2.28 & 564.47 & 6.98 & 1.34 \\ 
J0211+1051 & B & 0.2 & -12.0 & 7.66 & 1.6 & 12.31 & 3.92 & 2.08 & - & - & - \\ 
J0222+4302 & B & 0.34 & 240.33 & 3.72 & 3.22 & 27.0 & 1.5 & 3.0 & 400.56 & 29.91 & 1.31 \\ 
J0237+2848 & F & 1.213 & 120.17 & 3.61 & 2.4 & 6.8 & 4.6 & 2.65 & 493.19 & 1.95 & 1.67 \\ 
J0312+0133 & F & 0.664 & - & - & - & - & - & - & - & - & - \\ 
J0316+0904 & B & 0.372 & - & - & - & - & - & - & - & - & - \\ 
J0423-0120 & F & 0.915 & - & - & - & -0.48 & 5.17 & 3.08 & 118.48 & 6.07 & 2.27 \\ 
J0424+0036 & B & 0.266 & - & - & - & - & - & - & - & - & - \\ 
J0433+2905 & B & 0.97 & 24.69 & 0.0 & 1.0 & 27.01 & 0.0 & 1.33 & 10.47 & 3.66 & 1.88 \\ 
J0442-0017 & F & 0.844 & -66.75 & 4.03 & 1.93 & 11.81 & 4.32 & 2.1 & -549.1 & 1.67 & 1.17 \\ 
J0507+6737 & B & 0.314 & - & - & - & - & - & - & - & - & - \\ 
J0509+0541 & B & - & - & - & - & -24.65 & 4.76 & 1.19 & - & - & - \\ 
J0521+2112 & B & 0.108 & 98.61 & 11.09 & 2.66 & -33.25 & 2.51 & 3.37 & 244.93 & 11.95 & 2.78 \\ 
J0532+0732 & F & 1.254 & 77.72 & 3.4 & 1.82 & -4.87 & 0.65 & 2.57 & - & - & - \\ 
J0607+4739 & B & - & 293.17 & 4.58 & 1.52 & - & - & - & - & - & - \\ 
J0612+4122 & B & - & - & - & - & - & - & - & -75.37 & 2.71 & 2.05 \\ 
J0648+1516 & B & 0.179 & 108.86 & 0.0 & 2.47 & - & - & - & - & - & - \\ 
J0650+2502 & B & 0.203 & - & - & - & - & - & - & - & - & - \\ 
J0654+5042 & - & 1.253 & - & - & - & - & - & - & - & - & - \\ 
J0710+5908 & B & 0.125 & - & - & - & - & - & - & - & - & - \\ 
J0725+1425 & F & 1.038 & 229.15 & 7.5 & 1.51 & 8.94 & 0.0 & 1.78 & 563.85 & 3.32 & 1.2 \\ 
J0738+1742 & B & 0.424 & 51.16 & 3.93 & 2.65 & 2.03 & 10.78 & 3.47 & 96.62 & 18.67 & 1.45 \\ 
J0739+0137 & F & 0.191 & 62.76 & 3.09 & 1.3 & -0.89 & 6.62 & 2.0 & - & - & - \\ 
J0742+5444 & F & 0.723 & 110.97 & 3.09 & 2.35 & 19.9 & 4.23 & 3.11 & 155.58 & 3.66 & 2.54 \\ 
J0744+7433 & B & 0.314 & - & - & - & - & - & - & 328.12 & 0.0 & 1.19 \\ 
J0750+1231 & F & 0.889 & - & - & - & - & - & - & 199.55 & 16.57 & 1.4 \\ 
J0754-1147 & - & - & 188.55 & 0.13 & 2.14 & - & - & - & 300.39 & 0.0 & 1.78 \\ 
J0757+0956 & B & 0.266 & 577.3 & 4.43 & 1.46 & - & - & - & 278.81 & 12.45 & 1.5 \\ 
J0807-0541 & B & - & - & - & - & - & - & - & - & - & - \\ 
J0808-0751 & F & 1.837 & - & - & - & -9.35 & 0.14 & 1.82 & 119.27 & 7.21 & 2.47 \\ 
J0809+5218 & B & 0.137 & - & - & - & 51.48 & 3.48 & 3.15 & 166.79 & 5.2 & 1.22 \\ 
J0814+6431 & B & 0.239 & 249.79 & 3.65 & 2.05 & 9.28 & 1.52 & 2.5 & - & - & - \\ 
J0816-1311 & B & 0.046 & - & - & - & - & - & - & - & - & - \\ 
J0818+4222 & B & 0.53 & - & - & - & 1.47 & 6.35 & 2.0 & 217.02 & 4.87 & 1.81 \\ 
J0824+5552 & F & 1.417 & - & - & - & - & - & - & 236.28 & 6.15 & 1.0 \\ 
J0830+2410 & F & 0.94 & - & - & - & -50.99 & 4.12 & 3.02 & - & - & - \\ 
J0831+0429 & B & 0.174 & 609.16 & 4.79 & 1.23 & 7.23 & 10.21 & 1.31 & 678.41 & 5.78 & 1.29 \\ 
J0850-1213 & F & 0.566 & 70.03 & 4.5 & 1.72 & 22.97 & 4.39 & 2.0 & 91.59 & 15.0 & 2.03 \\ 
J0854+2006 & B & 0.306 & 187.16 & 4.8 & 1.57 & - & - & - & 591.64 & 8.9 & 1.9 \\ 
J0856-1105 & B & - & 150.5 & 6.88 & 1.44 & - & - & - & - & - & - \\ 
J0909+0121 & F & 1.024 & 95.94 & 8.12 & 3.1 & -27.5 & 3.53 & 3.1 & 453.03 & 15.91 & 1.53 \\ 
J0915+2933 & B & 1.521 & - & - & - & - & - & - & -2.03 & 0.0 & 2.0 \\ 
J0920+4441 & F & 2.19 & - & - & - & 6.55 & 16.43 & 1.1 & -152.03 & 8.46 & 1.75 \\
\hline
\end{tabular}
\end{table*}

\begin{table*}
\setlength{\tabcolsep}{10pt}
\centering
  \contcaption{}
\begin{tabular}{@{}llcccccccccc@{}}
 \hline
  (1) & (2) & (3) & (4) & (5) & (6) & (7) & (8) & (9) & (10) & (11) & (12) \\
  \hline
J0953-0840 & B & 0.59 & - & - & - & - & - & - & - & - & - \\ 
J0957+5522 & F & 0.899 & - & - & - & - & - & - & - & - & - \\ 
J1001+2911 & B & 0.08 & 250.82 & 12.07 & 2.05 & 7.82 & 23.42 & 2.0 & 241.14 & 6.28 & 2.5 \\ 
J1015+4926 & B & 1.19 & - & - & - & - & - & - & - & - & - \\ 
J1037+5711 & B & 0.831 & -26.89 & 2.28 & 2.25 & 2.84 & 2.4 & 3.1 & 416.11 & 1.86 & 1.41 \\ 
J1053+4929 & B & 0.14 & - & - & - & - & - & - & - & - & - \\ 
J1058+0133 & B & 0.185 & 56.7 & 45.89 & 2.14 & 15.33 & 4.28 & 2.5 & 95.51 & 7.05 & 1.1 \\ 
J1058+5628 & B & 0.143 & - & - & - & 14.35 & 11.19 & 2.63 & - & - & - \\ 
J1059-1134 & B & - & - & - & - & - & - & - & - & - & - \\ 
J1117+2014 & B & 0.138 & - & - & - & - & - & - & - & - & - \\ 
J1130-1449 & F & 1.189 & - & - & - & - & - & - & 483.28 & 15.9 & 2.56 \\ 
J1132+0034 & B & 0.678 & - & - & - & -13.44 & 0.0 & 2.17 & 146.04 & 0.0 & 1.61 \\ 
J1136+7009 & B & 0.046 & - & - & - & - & - & - & - & - & - \\ 
J1150+4154 & B & 0.004 & - & - & - & - & - & - & - & - & - \\ 
J1159+2914 & F & 0.729 & - & - & - & 30.57 & 10.2 & 1.6 & 89.83 & 6.67 & 1.65 \\ 
J1217+3007 & B & 0.13 & 212.86 & 1.6 & 3.0 & - & - & - & 363.53 & 5.77 & 1.75 \\ 
J1221+2813 & B & 0.102 & - & - & - & -8.46 & 7.79 & 1.6 & - & - & - \\ 
J1221+3010 & B & 0.184 & - & - & - & - & - & - & - & - & - \\ 
J1222+0413 & F & 0.965 & 13.53 & 74.7 & 1.06 & - & - & - & 177.49 & 3.43 & 1.7 \\ 
J1224+2122 & F & 0.434 & 58.38 & 17.18 & 2.1 & -23.41 & 7.13 & 2.12 & - & - & - \\ 
J1231+2847 & B & 0.236 & - & - & - & - & - & - & - & - & - \\ 
J1243+3627 & B & 1.066 & - & - & - & - & - & - & - & - & - \\ 
J1248+5820 & B & 0.847 & 156.83 & 4.1 & 1.62 & 28.91 & 8.73 & 2.11 & - & - & - \\ 
J1253+5301 & B & 1.084 & - & - & - & 13.51 & 8.95 & 2.21 & - & - & - \\ 
J1256-0547 & F & 0.536 & 31.38 & 3.96 & 2.37 & -49.95 & 11.5 & 1.9 & - & - & - \\ 
J1309+4305 & B & 0.691 & - & - & - & - & - & - & - & - & - \\ 
J1310+3220 & F & 0.997 & -12.83 & 2.23 & 1.26 & 237.56 & 10.7 & 1.44 & - & - & - \\ 
J1314+2348 & B & 1.54 & 103.56 & 4.19 & 2.63 & -43.35 & 7.81 & 3.37 & - & - & - \\ 
J1337-1257 & F & 0.539 & - & - & - & - & - & - & -329.91 & 4.31 & 1.65 \\ 
J1351+1114 & B & 0.395 & - & - & - & - & - & - & - & - & - \\ 
J1354-1041 & F & 0.332 & - & - & - & 31.53 & 3.7 & 1.9 & 679.99 & 0.94 & 1.33 \\ 
J1418-0233 & B & 0.0 & 15.13 & 7.4 & 2.45 & 55.1 & 14.22 & 3.17 & 161.88 & 15.15 & 1.57 \\ 
J1427+2347 & B & 0.16 & - & - & - & -33.43 & 7.06 & 2.28 & - & - & - \\ 
J1428+4240 & B & 0.129 & - & - & - & - & - & - & - & - & - \\ 
J1436+5639 & B & 0.17 & - & - & - & - & - & - & - & - & - \\ 
J1440+0610 & B & 0.396 & - & - & - & - & - & - & 67.68 & 10.41 & 2.3 \\ 
J1448+3608 & B & 1.508 & - & - & - & - & - & - & - & - & - \\ 
J1501+2238 & B & 0.235 & 71.02 & 0.0 & 2.01 & - & - & - & 437.02 & 5.37 & 1.33 \\ 
J1505+0326 & - & - & - & - & - & - & - & - & 91.83 & 16.57 & 2.25 \\ 
J1542+6129 & B & 0.507 & - & - & - & -0.99 & 2.54 & 2.37 & - & - & - \\ 
J1549+0237 & F & 0.414 & - & - & - & - & - & - & - & - & - \\ 
J1553+1256 & F & 1.308 & - & - & - & - & - & - & - & - & - \\ 
J1555+1111 & B & 0.36 & 127.48 & 1.84 & 1.44 & 10.03 & 12.38 & 2.04 & 93.89 & 25.53 & 1.94 \\ 
J1558+5625 & B & 0.3 & - & - & - & - & - & - & 107.4 & 5.58 & 1.46 \\ 
J1608+1029 & F & 1.226 & - & - & - & - & - & - & 68.85 & 1.88 & 1.51 \\ 
J1635+3808 & F & 1.814 & 18.95 & 5.92 & 1.46 & -21.26 & 6.72 & 2.16 & 52.61 & 2.05 & 2.41 \\ 
J1642+3948 & F & 0.593 & 26.13 & 1.92 & 2.04 & 85.37 & 8.46 & 1.0 & -304.31 & 2.75 & 1.36 \\ 
J1643-0646 & - & 0.082 & - & - & - & - & - & - & -323.73 & 0.65 & 1.71 \\ 
J1719+1745 & B & 0.137 & - & - & - & 19.25 & 13.59 & 2.0 & - & - & - \\ 
J1725+1152 & B & 0.018 & - & - & - & - & - & - & 239.21 & 0.0 & 1.02 \\ 
J1725+5851 & B & 0.001 & - & - & - & - & - & - & 269.73 & 0.0 & 3.12 \\ 
J1727+4530 & F & 0.714 & - & - & - & - & - & - & 255.15 & 13.38 & 1.52 \\ 
J1740+5211 & F & 1.379 & - & - & - & 186.11 & 1.7 & 1.78 & - & - & - \\ 
J1748+7005 & B & 0.77 & 426.14 & 3.13 & 1.0 & 14.64 & 7.97 & 3.23 & 88.21 & 25.94 & 1.23 \\ 
J1751+0939 & B & 0.322 & - & - & - & - & - & - & - & - & - \\ 
J1754+3212 & B & 1.09 & - & - & - & 12.14 & 7.2 & 3.0 & 163.12 & 9.59 & 2.46 \\ 
J1800+7828 & B & 0.68 & - & - & - & 4.79 & 6.66 & 3.02 & 96.26 & 8.75 & 1.0 \\ 
J1813+3144 & B & 0.117 & - & - & - & - & - & - & - & - & - \\ 
J1824+5651 & B & 0.664 & - & - & - & 7.65 & 5.9 & 2.13 & - & - & - \\  
\hline
\end{tabular}
\end{table*}
\begin{table*}
\setlength{\tabcolsep}{10pt}
\centering
  \contcaption{}
\begin{tabular}{@{}llcccccccccc@{}}
 \hline
  (1) & (2) & (3) & (4) & (5) & (6) & (7) & (8) & (9) & (10) & (11) & (12) \\
  \hline
J1849+6705 & F & 0.657 & 30.33 & 2.96 & 1.22 & -3.86 & 4.85 & 2.68 & 40.74 & 15.67 & 1.0 \\ 
J1903+5540 & B & 0.58 & 66.45 & 1.02 & 1.54 & 5.03 & 3.51 & 2.47 & 161.91 & 10.31 & 1.74 \\ 
J1917-1921 & B & 0.137 & 289.78 & 3.05 & 4.9 & 39.54 & 21.9 & 1.57 & -8.29 & 5.9 & 1.6 \\ 
J1921-1607 & B & - & - & - & - & - & - & - & - & - & - \\ 
J1926+6154 & B & - & - & - & - & - & - & - & - & - & - \\ 
J1959+6508 & B & 0.049 & -51.91 & 0.56 & 3.21 & -5.49 & 6.09 & 1.01 & - & - & - \\ 
J2000-1748 & F & 0.652 & 12.82 & 2.35 & 2.05 & - & - & - & -13.05 & 3.76 & 2.03 \\ 
J2012+4628 & B & - & 466.61 & 1.19 & 1.36 & 7.35 & 6.44 & 3.03 & 511.51 & 4.84 & 2.45 \\ 
J2039-1046 & B & 1.05 & 221.13 & 0.87 & 2.6 & - & - & - & - & - & - \\ 
J2055-0021 & B & 0.407 & - & - & - & - & - & - & - & - & - \\ 
J2116+3339 & B & 1.596 & - & - & - & - & - & - & 264.15 & 16.91 & 3.01 \\ 
J2143+1743 & F & 0.211 & - & - & - & - & - & - & 615.35 & 18.14 & 1.55 \\ 
J2147+0929 & F & 1.113 & 170.1 & 2.99 & 1.75 & 17.3 & 0.0 & 1.0 & - & - & - \\ 
J2152+1734 & B & 0.871 & - & - & - & - & - & - & - & - & - \\ 
J2202+4216 & B & 0.069 & - & - & - & -7.78 & 5.58 & 2.01 & - & - & - \\ 
J2203+1725 & F & 1.076 & - & - & - & - & - & - & 553.88 & 3.27 & 2.16 \\ 
J2225-0457 & F & 1.404 & 255.3 & 0.05 & 1.01 & - & - & - & 225.48 & 7.48 & 2.27 \\ 
J2229-0832 & F & 1.559 & 239.34 & 5.44 & 1.2 & 25.84 & 9.03 & 1.0 & 62.09 & 15.75 & 1.74 \\ 
J2232+1143 & F & 1.037 & 300.3 & 0.3 & 2.17 & 24.49 & 1.32 & 3.12 & - & - & - \\ 
J2236-1433 & B & 0.325 & 718.5 &  2.74 & 1.71 & 25.27 & 8.0 & 3.26 & 715.5 & 1.35 & 1.12 \\ 
J2236+2828 & F & 0.795 & 11.73 & 3.14 & 1.28 & -2.11 & 15.03 & 2.5 & -4.72 & 5.43 & 1.01 \\ 
J2243+2021 & B & 0.39 & - & - & - & 78.26 & 14.17 & 1.34 & - & - & - \\ 
J2250+3824 & B & 0.119 & - & - & - & - & - & - & - & - & - \\ 
J2251+4030 & B & 0.229 & - & - & - & - & - & - & - & - & - \\ 
J2253+1608 & F & 0.859 & 402.95 & 5.93 & 2.36 & -5.72 & 2.77 & 2.57 & 115.49 & 5.85 & 2.5 \\ 
J2323+4210 & B & 0.059 & - & - & - & - & - & - & - & - & - \\ 
J2329+3754 & - & 0.264 & - & - & - & - & - & - & - & - & - \\ 
J2345-1555 & F & 0.621 & 20.58 & 0.97 & 2.38 & -11.58 & 0.0 & 3.36 & 93.82 & 1.32 & 2.06 \\ 
J2347+5142 & B & 0.044 & 57.37 & 0.0 & 1.35 & -13.36 & 28.62 & 1.01 & 100.96 & 1.57 & 2.04 \\ 
J2348-1631 & F & 0.576 & - & - & - & - & - & - & - & - & - \\ 
\hline  
\end{tabular}
\end{table*}

\label{lastpage}
\end{document}